\documentclass[showpacs,showkeys]{revtex4}
\usepackage{graphicx}

\usepackage{amsmath}
\usepackage{amssymb}
\usepackage{array}
\usepackage{color}

\begin{document}

\title{Flux tube in turbulent flow and quantum chromodynamics}

\author{Vladimir Dzhunushaliev}
\affiliation{Institut f\"ur  Physik, Universit\"at Oldenburg, Postfach 2503, 
D-26111 Oldenburg, Germany; 
\\
Department of Physics and Microelectronic
Engineering, Kyrgyz-Russian Slavic University, Bishkek, Kievskaya Str.
44, 720021, Kyrgyz Republic \\
and \\
Institute of Physics of National Academy of Science
Kyrgyz Republic, 265 a, Chui Street, Bishkek, 720071,  Kyrgyz Republic}
\email[Email: ]{vdzhunus@krsu.edu.kg}

\date{\today}

\begin{abstract}
Using some assumptions about the correlation function of velocity in a turbulent flow, a cylindrically symmetric
tube-like solution is obtained. It is proposed that this turbulent flow is similar to a flux tube in quantum
chromodynamics. Using similar assumptions for quantum chromodynamics, a flux tube filled with a color electric
field is obtained. The connection between turbulence theory and quantum chromodynamics is discussed.
\end{abstract}


\maketitle

\section{Introduction}

Two of the most serious problems in modern physics are the creation of turbulence and the quantization of strongly interacting quantum fields (nonperturbative quantization). From Ref. \cite{Energy}: "In the world of physics, the difficulty of understanding turbulence in fluids is legendary. A famous quotation is also variously attributed to Einstein, Heisenberg, Richard Feynman, or Arnold Sommerfeld: �Turbulence is the last great unsolved problem of classical physics." In a similar way the nonperturbative quantization is unsolved problems in quantum physics. The nonperturbative quantization problem is connected with the problem of quark confinement or unobservability of a single quark. From Wikipedia \cite{wiki}: ``Color confinement is the physics phenomenon that color charged particles (such as quarks) cannot be isolated singularly, and therefore cannot be directly observed. The reasons for quark confinement are somewhat complicated; there is no analytic proof that quantum chromodynamics should be confining, but intuitively, confinement is due to the force-carrying gluons having color charge. As any two electrically-charged particles separate, the electric fields between them diminish quickly, allowing (for example) electrons to become unbound from nuclei. However, as two quarks separate, the gluon fields form narrow tubes (or strings) of color charge, which tend to bring the quarks together as though they were some kind of rubber band. This is quite different in behavior from electrical charge. Because of this behavior, the color force experienced by the quarks in the direction to hold them together, remains constant, regardless of their distance from each other.'' Here we would like to consider a possible connection between these two problems. 

In order to understand the reason of confinement problem in quantum chromodynamics there are many attempts to compare quantum chromodynamics witn other branches of physics. In Ref.~\cite{hooft} it was suggested that the confinement of quarks into hadrons may happen due to a condensation of special gluonic configurations called Abelian monopoles. In this approach (the dual superconductivity scenario) a condensate of the monopoles breaks spontaneously an internal $U(1)$ gauge symmetry. According to the dual superconductor idea, the breaking of the dual symmetry gives rise naturally to the dual Meissner effect, which insures a formation of a tube, which in turn leads to the confinement the quarks into hadronic bound states. In Ref.~\cite{Chernodub:2005jh} it is supposed that the $SU(2)$ Yang-Mills theory in the low-temperature phase
can be considered as a nematic liquid crystal. In Ref. \cite{niemi} it is proposed that the SU(2) Yang-Mills theory can be interpreted as a two-band dual superconductor with an interband Josephson coupling. This approach is based on a slave-boson decomposition \cite{highTc} used in the high-T$_c$ superconductivity models (similar decomposition used in gauge theories is named as a spin-charge separation \cite{niemi} \cite{faddeev}) . 

The color "magnetic" flux tubes which appear in QCD with high baryon density in Ref's \cite{Balachandran:2005ev}-\cite{Eto:2009kg} are considered. QCD with high density at low temperature is in the color-flavor locked(CFL) phase, where it exhibits color superconductivity. Unlike color electric/magnetic fluxes in gluon plasma, those color magnetic flux tubes are topologically and dynamically stable. They are a mixture of a superconductor vortex and a superfluid vortex and are called a semi-superfluid vortices. 

The color magnetic flux tubes also exist (as BPS vortices) in supersymmetric QCD. In Ref's \cite{Tong:2005un}-\cite{Tong:2008qd} one can find reviews on this subject. 

\section{Reynolds number and coupling constant}

Why such connection between turbulence and quantum chromodynamics may exists ? In both cases we have the fluctuations: velocity, pressure and so on for a turbulent fluid in the first case and fields (for example, SU(3) gauge fields for quantum chromodynamics) in the second case. In the turbulent fluid the fluctuations are statistical ones and in quantum field theory the fluctuations are quantum ones. 

The character of fluid flow depends on Reynolds number $\mathrm{Re}$
\begin{equation}
	\mathrm{Re} = \frac{\rho_m v l}{\mu}
\label{1-10}
\end{equation}
where $\rho_m$ is the fluid density, $v$ is the fluid velocity, $l$ is a characteristic length for a given flow and $\mu$ is the fluid viscosity. If $\mathrm{Re} < \mathrm{Re}_{cr}$ then the flow is laminar one, if $\mathrm{Re} > \mathrm{Re}_{cr}$ then the flow is turbulent one. The open question in hydrodynamics is the creation of turbulence by 
$\mathrm{Re} \approx \mathrm{Re}_{cr}$. 

In quantum field theory there are perturbative regime when a dimensionless coupling constant 
\begin{equation}
	\alpha^2 = \frac{1/\tilde g^2}{\hbar c}
\label{1-20}
\end{equation}
is small enough $\alpha^2 < 1$ and nonperturbative regime when $\alpha^2 \geq 1$ (here $\tilde g$ is a dimension coupling constant, $\hbar$ is Planck constant and $c$ is the speed of light). In quantum field theory the open question is a nonperturbative quantization for $\alpha^2 \geq 1$. For example, the fine-structure constant in quantum electrodynamics is 
$\alpha^2 = e^2/\hbar c \approx 1/137$ ($e$ is the electron charge), $\alpha^2 \approx 0.1$ for a weak interaction and $\alpha^2 \geq 1$ for a strong interaction. 

Here we propose a possible connection between hydrodynamics and quantum field theory based on a comparison between Reynolds number (in hydrodynamics) and dimensionless coupling constant (in quantum field theory). Let us rewrite Eq. \eqref{1-10} in following form 
\begin{equation}
	\mathrm{Re} = \frac{\rho v^2 l^4}{\mu l^3 v}
\label{1-30}
\end{equation}
in such a way that 
$\left[ \rho v^2 l^4 \right] = \left[ 1/\tilde g^2 \right] = g \cdot cm^3 / s^2$, 
$\left[ \hbar \right] = \left[ \mu l^3 \right] = g \cdot cm^3 / s$. Now we will assume that there is following relation 
\begin{eqnarray}
	1/\tilde g^2 & \leftrightarrow & \rho v^2 l^4 ,
\label{1-40}\\
	\hbar & \leftrightarrow & \mu l^3 .
\label{1-50}
\end{eqnarray}
It means that we have following correspondence 
\begin{itemize}
	\item $\hbar = 0 \leftrightarrow \mu = 0$. In this case a classical theory corresponds to an ideal fluid.
	\item $\hbar \neq 0 \text{ and } \alpha^2 < 1 \leftrightarrow \mu \neq 0 \text{ and } \mathrm{Re} < \mathrm{Re}_{cr}$. In this case a laminar fluid corresponds to a perturbative regime of quantum field theory.
	\item $\hbar \neq 0 \text{ and } \alpha^2 \geq 1 \leftrightarrow \mu \neq 0 \text{ and } \mathrm{Re} > \mathrm{Re}_{cr}$. In this case a turbulent fluid corresponds to a nonperturbative regime of quantum field theory.
\end{itemize}

\section{Flux tube in turbulent flow}
\label{turbulence}

How we can use proposed here the similarity between hydrodynamics ans nonperturbative quantum field theory ? One can try to find in hydrodynamics an analog of flux tube from  quantum chromodynamics. 

At first we will try to examine a static ideal and laminar fluids. The Navier - Stokes equations are 
\begin{equation}
	\frac{\partial \vec v}{\partial t} + \left(
		\vec v \, \nabla
	\right) \vec v = - \frac{1}{\rho_m} \nabla p + 
	\nu \Delta \vec v 
\label{2-10}
\end{equation}
here we consider incompressible fluid $\mathrm {div} \vec v = 0$ and $\nu = \frac{\eta}{\rho_m}$ is the kinematic viscosity. If the viscosity $\mu=0$ then we have the ideal fluid. We want to find cylindrically symmetric solution of the Navier - Stokes equations but not in a pipe. For the ideal $(\nu=0)$ and laminar fluids the equation is 
\begin{equation}
	v_\rho \frac{d v_z}{d \rho} = \nu \frac{d^2 v_z}{d \rho^2}
\label{2-15}
\end{equation}
but $v_\rho = 0$ for the flow along axes $z$. Consequently the solution for the ideal fluid does not exist. For the laminar fluid we have $d^2 v_z/d\rho^2 = 0$ and this equation has a solution in a pipe only. Only one possibility exists to have a nontrivial solution: it is necessary to have the correlation between $v_\rho$ and $d v_z/d \rho$ in the LHS of Eq. \eqref{2-15}. 

Now we would like to consider a turbulent fluid. We will try to find an axially symmetric solutions. We consider the fluid in unbounded space with two sources located at the $z = \pm \infty$. Then the equations \eqref{2-10} are 
\begin{eqnarray}
	\nu \frac{1}{\rho} \frac{d }{d \rho} \left( \rho v_z \right)  &=& 
	v_z \frac{\partial v_z}{\partial z} + 
	v_\rho \frac{\partial v_z}{\partial \rho} + 
	v_\phi \frac{\partial v_z}{\partial \phi} ,
\label{2-30}\\
	\nu \frac{1}{\rho} \frac{d }{d \rho} \left( \rho v_\rho \right)  &=& 
	v_z \frac{\partial v_\rho}{\partial z} + 
	v_\rho \frac{\partial v_\rho}{\partial \rho} + 
	v_\phi \frac{\partial v_\rho}{\partial \phi} ,
\label{2-40}\\
	\nu \frac{1}{\rho} \frac{d }{d \rho} \left( \rho v_\phi \right)  &=& 
	v_z \frac{\partial v_\phi}{\partial z} + 
	v_\rho \frac{\partial v_\phi}{\partial \rho} + 
	v_\phi \frac{\partial v_\phi}{\partial \phi} .
\label{2-50}
\end{eqnarray}
We assume that in the first approximation such flow can be described with fluctuating velocities $v_{z, \rho}$ that depend on $\rho$ only. We use the cylindrical coordinate system $z, r, \phi$.  The flow is along the axes $z$: $\bar{v}_z \neq 0$ and $\bar v_\rho = 0$, $\overline{(\cdots)}$ means the statistical overaging. Remember that we consider the solutions in unbounded space, not in a pipe.

Now we can write equation for $v_z$ velocity 
\begin{equation}
	\nu \frac{1}{\rho} \frac{d }{d \rho} \left( \rho v_z \right) = v_\rho \frac{d v_z}{d \rho} .
\label{2-60}
\end{equation}
Unfortunately it is not too easy to write equation for $v_\rho$ velocity. For example, if we write equation in the form  
\begin{equation}
	\nu \frac{1}{\rho} \frac{d }{d \rho} \left( \rho v_\rho \right) = 
	v_\rho \frac{d v_\rho}{d \rho} .
\label{2-70}
\end{equation}
and after averaging we will have the correlation 
\begin{equation}
	\overline{v_\rho \frac{d v_\rho}{d \rho}} = 0 
\label{2-80}
\end{equation}
because $\bar v_\rho = 0$. It happens because the terms $v_z \frac{\partial v_\rho}{\partial z}$ and 
$v_\phi \frac{\partial v_\rho}{\partial \phi}$ are not taking into account in Eq. \eqref{2-40}. 

In order to write equation for $v_\rho$ we multiply Eq. \eqref{2-40} on $v_\rho$ and only after that we average over the statistical fluctuations. In this case we assume that one can neglect with the terms $v_z \frac{\partial v_\rho}{\partial z}$ and $v_\phi \frac{\partial v_\rho}{\partial \phi}$ 
\begin{equation}
	\nu v_\rho v''_\rho + 
	\frac{\nu}{\rho} v_\rho  v'_\rho = 
	v^2_\rho v'_\rho
\label{2-90}
\end{equation}
where $(\cdots)' = d(\cdots) / d \rho$. After some manipulations and averaging we have
\begin{equation}
	\frac{\nu}{2} \left (\overline{v_\rho^2} \right )'' - 
	\nu \overline{\left (v'_\rho \right )^2} + 
	\frac{\nu}{2 \rho} \overline{\left (v_\rho^2 \right )'} = 
	\frac{1}{3} \overline{v^2_\rho v'_\rho}\,
\label{2-100}
\end{equation} 
here we neglect with $\partial / \partial z$ and $\partial / \partial \phi$ terms in Eq. \eqref{2-40}. Now we have to do some assumptions about correlations functions in \eqref{2-100}. We should have the information about 2 and 3 points correlation functions. The most important simplification is that all correlation fuctions are described by one function $\phi(x^i), i=1,2,3$: 
\begin{eqnarray}
	\overline{v_a(x^i) v_b(y^i)} &=& \delta_{ab} \phi(x^i) \phi(y^i) , 
\label{2-110}\\
	\overline{v_a(x^i) v_b(y^i) \frac{\partial v_c(z^i)}{\partial z^i}} &=& 
	\left( A_{abc} \phi(x^i) \phi(y^i) \frac{\partial \phi(z^i)}{\partial z^i} + \text{ transmutation by } (x,y,z) \right) +  
\nonumber \\
	&&
	\left( \alpha_{a} \phi(y^i) \phi(z^i) + 
	\text{ transmutation by } (x,y,z) \right) + 
\nonumber \\
	&&
	\left( \beta_{a} \phi(y^i) \frac{\partial \phi(z^i)}{\partial z^i} + 
	\text{ transmutation by } (x,y,z) \right)
\label{2-120}
\end{eqnarray}
here $A_{abc}$ and $\alpha_{a}, \beta_a$ are some constants, $a,b,c = \rho, \phi$. The correlation function \eqref{2-120} is choosen in such way because we assume that $\bar v_{a} = 0$ and conjectural corrections to the first term $\phi(x^i) \phi(y^i) \frac{\partial \phi(z^i)}{\partial z^i}$ can be only quadratic compositions: either $\phi(y^i) \phi(z^i)$ or $\phi(y^i) \frac{\partial \phi(z^i)}{\partial z^i}$. For $v_z$ we have to use not $v_z$ but: 
$\tilde v_z = v_z - \bar v_z$ 
\begin{eqnarray}
	\overline{v_a(x^i) \tilde v_z(y^i)} &=& \gamma \phi(x^i) \phi(y^i) , 
\label{2-130}\\
	\overline{v_a(x^i) v_b(y^i) \tilde v_z(z^i)} &=& \gamma \left[ 
	\left( A_{abc} \phi(x^i) \phi(y^i)\phi(z^i) + \text{ transmutation by } (x,y,z) \right) + 
	\right. 
\nonumber \\
	&& \left. 
	\left( \alpha_{a} \phi(y^i) \phi(z^i) + \text{ transmutation by } (x,y,z) \right) 
	\right] 
\label{2-150}
\end{eqnarray}
where $\gamma$ is a constant indicating the difference in the scale between longitudinal velocity $v_z$ and the fluctuations of transversal velocity $v_{\rho}$. 

\begin{figure}[h]
	\includegraphics[width=12cm]{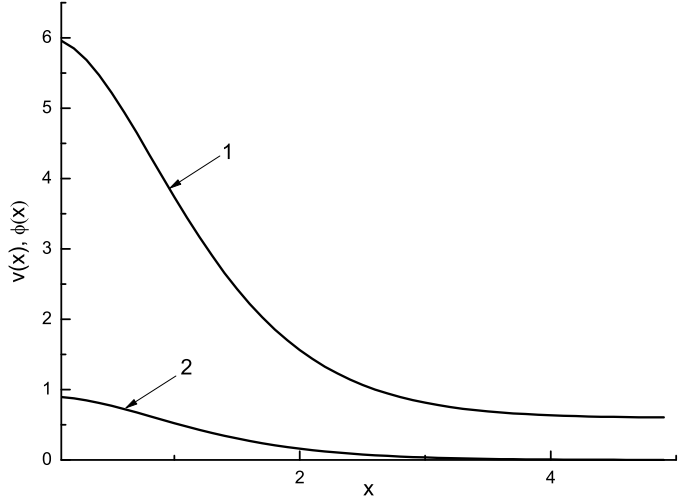}
	\caption{The profiles of functions  $v(x)$ and $\phi(x)$, 
	$\tilde \alpha = \tilde \beta = - 3, \gamma = 6$. $v(x)$ - curve 1, $\phi(x)$ - curve 2} 
	\label{fig1}
\end{figure}

According to \eqref{2-110} \eqref{2-120}
\begin{eqnarray}
	\overline{v_\rho^2} &=& \phi^2 ,
\label{2-160}\\
	\overline{\left( {v'_\rho}^2 \right)} &=& \lim_{\substack{\rho_2 \rightarrow \rho_1}} 
	\frac{d}{d \rho_1} \frac{d}{d \rho_2} 
	\left( 
	\overline{v_\rho(\rho_1) v_\rho(\rho_2)}
	\right) = \phi'^2 , 
\label{2-170}\\
	\left( 
	  \overline{v^2_\rho v'_\rho} 
	\right)' &=& 3 \phi^2 \phi' + 2 \alpha \phi^2 + 2 \beta \phi \phi'. 
\label{2-180}
\end{eqnarray} 
Thus we have following equation 
\begin{equation}
	\nu \phi'' + \frac{\nu}{\rho} \phi' = 
	\phi \phi' + \frac{2}{3} \alpha \phi + \frac{2}{3} \beta \phi' . 
\label{2-190}
\end{equation} 
Now we would like to write equation for $\bar v_z  = v_z + \delta v_z$ velocity component. After substitution into Eq.  \eqref{2-60}, averaging over statistic fluctuation we have 
\begin{equation}
	\nu \bar v_z'' + \frac{\nu}{\rho} \bar v_z' = \overline{ v_\rho \delta v_z'} 
\label{2-200}
\end{equation}  
here the RHS is similar to the RHS of Eq. \eqref{2-70} and consequently similar calculations leads to 
\begin{equation}
	\nu \bar v_z'' + \frac{\nu}{\rho} \bar v_z' = 
	\gamma \left( \phi \phi' + \frac{2}{3} \alpha \phi + \frac{2}{3} \beta \phi' \right) 
\label{2-210}
\end{equation} 
here the number $\gamma$ appears in the consequence of correlations \eqref{2-130} \eqref{2-150}. Thus we have following set of equations for the longitudinal velocity component $\bar v_z$ and fluctuating transversal velocity component $v_\rho$ 
\begin{eqnarray}
	\tilde \phi'' + \frac{1}{x} \tilde \phi' &=& 
	\tilde \phi \tilde \phi' + \tilde \alpha \tilde \phi + \tilde \beta \phi'  ,
\label{2-220}\\
	v'' + \frac{1}{x} v' &=& 
	\gamma \left( \tilde \phi \tilde \phi' + \tilde \alpha \tilde \phi + 
\tilde \beta \tilde \phi' \right) 
\label{2-230}
\end{eqnarray} 
where we have introduced dimensionless variables $x=\rho/\rho_0$, $\rho_0 = \nu/\phi_0$, $\tilde \phi=\phi/\phi_0$, 
$\tilde \alpha = \frac{2}{3} \alpha \nu/\phi_0^2$, $\tilde \beta = \frac{2}{3} \beta / \phi_o$, $v = \bar v_z / \phi_0$ and $\phi_0 = \phi(0)$. The numerical solution of the set \eqref{2-220} \eqref{2-230} is presented in Fig. \ref{fig1}. Let us note that the solution exists only for negative values $\alpha < 0, \beta < 0$. 

Our interpretation of this solution is follows. There exists a turbulent flow moving in $z$ direction. In the flow exists a core (tube) with the velocity profile given in Fig. \ref{fig1}, curve 1; the $z-$component of the velocity at the infinity is $v_\infty \neq 0, \rho \rightarrow \infty$; the profile of statistical fluctuations describing by the function $\phi$ is given by curve 2 in Fig. \ref{fig1}; as $\phi \rightarrow 0$ then at the infinity the flow become laminar one. It means that we have to choose the solution in such a way that $v_\infty$ becomes small enough that flow becomes laminar one. In our opinion it is an analog of a flux tube filled with a gauge fields and stretched between quark and antiquark from quantum chromodynamics. 

Such kind of solution is characterized by having a strongly pronounced core (tube) through which flows the turbulent fluid. Such flux of the turbulent fluid may be inserted either into a flow of a laminar fluid $v_\infty \neq 0$ (that is more probably) or in a rest fluid 
$v_\infty = 0$ (that is less probably). 

In summary we list the basic ideas and assumptions in presented approximation:
\begin{itemize}
	\item All degrees of freedom are splitted into two parts: ordered movement 
	$\bar v_z$ and disordered fluctuations $v_\rho$. 
	\item The correlation functions of the second and third orders are expressed in terms of one function (one-function approximation). 
	\item The correlation function of the third order is expressed in terms of 
the correlation function of the second order. 
\end{itemize}
The last assumption allow us to cut off an infinite equations set relating all correlation functions of all orders. 

\section{Flux tube in quantum chromodynamics}

In this section we would like to show that similar calculations can be done in quantum chromodynamics. The basic assumptions will be same as listed at the end of Section \ref{turbulence}. 


In this section we follow the conventions of Ref. \cite{kondo}. Starting with the $SU(N)$ gauge group with generators $T^B$ we define the $SU(N)$ gauge fields, $\mathcal{A}_\mu=\mathcal{A}^B_\mu T^B$. Let $G$ be a subgroup of $SU(N)$ and $SU(N)/G$ is a coset. Then the gauge field $\mathcal{A}_\mu$ can be decomposed as
\begin{eqnarray}
  \mathcal{A}_\mu & = & \mathcal{A}^B_\mu T^B = a^a_\mu T^a + A^m_\mu T^m ,
\label{sec3-10}\\
  a^a_\mu & \in & G \quad \text{and} \quad A^m_\mu \in SU(N)/G
\label{sec3-20}
\end{eqnarray}
where the indices $a,b,c \ldots $ belongs to the subgroup G and 
$m,n, \ldots $to the coset $SU(N)/G$; $B$ are $SU(N)$ indices. Based on this the field strength can be decomposed as
\begin{equation}
  \mathcal{F}^B_{\mu\nu} T^B = \mathcal{F}^a_{\mu\nu}T^a +
  \mathcal{F}^m_{\mu\nu}T^m
\label{sec3-30}
\end{equation}
where 
\begin{eqnarray}
  \mathcal{F}^a_{\mu\nu} & = & \phi ^a_{\mu\nu} + \Phi^a_{\mu\nu}
  \; \; \; \in G ,
\label{sec3-40}\\
  \phi ^a_{\mu\nu} & = & \partial_\mu a^a_\nu - \partial_\nu a^a_\mu +
  f^{abc}a^b_\mu a^c_\nu \; \; \; \in G ,
\label{sec3-50}\\
  \Phi^a_{\mu\nu} & = & f^{amn} A^m_\mu A^n_\nu \; \; \; \in G ,
\label{sec3-60}\\
  \mathcal{F}^m_{\mu\nu} & = & F^m_{\mu\nu} + G^m_{\mu\nu} \; \; \;
  \in SU(N)/G ,
\label{sec3-70}\\
  F^m_{\mu\nu} & = & \partial_\mu A^m_\nu - \partial_\nu A^m_\mu +
  f^{mnp} A^n_\mu A^p_\nu \; \; \; \in SU(N)/G ,
\label{sec3-80}\\
  G^m_{\mu\nu} & = & f^{mnb}
  \left(
  A^n_\mu a^b_\nu - A^n_\nu a^b_\mu
  \right) \; \; \; \in SU(N)/G
\label{sec3-90}
\end{eqnarray}
where $f^{ABC}$ are the structural constants of $SU(N)$. The $SU(N)$ Yang-Mills Lagrangian can be decomposed as
\begin{equation}
  \mathcal L = \mathcal{F}^B_{\mu\nu} \mathcal{F}_{B \mu\nu} = 
  \mathcal{F}^a_{\mu\nu} \mathcal{F}_{a \mu\nu} + 
  \mathcal{F}^m_{\mu\nu} \mathcal{F}_{m \mu\nu}.
\label{sec3-110}
\end{equation}
We will specialize to: 
\begin{itemize}
	\item the $SU(2)$ case when we let $SU(N) \rightarrow SU(2)$, 
	$G \rightarrow U(1)$, and $f^{ABC} \rightarrow \epsilon ^{ABC}$; the indices $a = 3, m,n = 1,2$ (here $\epsilon^{ABC}=0,\pm 1$ is the absolutely antisymmetric Levi-Civita tensor).
	\item the $SU(3)$ case when we let $SU(N) \rightarrow SU(3)$, 
	$G \rightarrow SU(2)$; the indices $a = 1,2,3, m,n = 4,5,6,7,8$.
\end{itemize}
For the first case one can assume that the fourth order Green function can be decomposed as the product of the second order Green functions and obtain Ginzburg-Landau Lagrangian \cite{Dzhunushaliev:2002xr}. 

For our purposes the second case is more interesting: using all assumptions from Section \ref{turbulence} with some modifications we will show that in SU(3) gauge theory exists a flux tube. In this case a longitudinal color electric field $E^3_z$ will be analog of the velocity $\bar v_z$ and gauge potentials $A^m_\mu$ in the coset $SU(3)/SU(2)$ are the analog of fluctuating transversal velocity $v_\rho$.

In quantizing the classical field equations system (via Heisenberg's non-perturbative method \cite{heisenberg}) one first replaces the classical fields by field operators 
$\mathcal A^B_{\mu} \rightarrow \widehat{\mathcal A}^B_\mu$. This yields the
following Yang-Mills equations for the operators
\begin{equation}
    \partial_\nu \widehat {\mathcal F}^{B\mu\nu} = 0.
\label{sec3-120}
\end{equation}
These nonlinear equations for the field operators of the nonlinear quantum fields can be used to determine expectation values for the field operators
$\widehat {\mathcal A}^B_\mu$, where 
$\langle \cdots \rangle = \langle Q | \cdots | Q \rangle$ and
$| Q \rangle$ is some quantum state. One can also use these equations to determine the expectation values of operators that are built up from the fundamental operators $\widehat {\mathcal A}^B_\mu$. For example, the ``electric'' field operator, 
$\widehat {\mathcal E}^B_z = \partial _0 \widehat {\mathcal A}^B_z -
\partial _z \widehat {\mathcal A}^B_0 + g f^{BCD} \mathcal A^C_0 \mathcal A^D_z$ 
giving the expectation $\langle \widehat {\mathcal E}^B_z \rangle$.
The simple gauge field expectation values, 
$\langle \mathcal{A}_\mu (x) \rangle$, are obtained by average Eq. \eqref{sec3-120} over some quantum state $| Q \rangle$
\begin{equation}
  \left\langle Q \left|
  \partial_\nu \widehat {\mathcal F}^{B\mu\nu}
  \right| Q \right\rangle = 0.
\label{sec3-130}
\end{equation}
One problem in using these equations to obtain expectation values like 
$\langle \mathcal A^B_\mu \rangle$, is that these equations involve not only powers or derivatives of $\langle \mathcal A^B_\mu \rangle$ ({\it i.e.} terms like 
$\partial_\alpha \langle \mathcal A^B_\mu \rangle$ or $\partial_\alpha
\partial_\beta \langle \mathcal A^B_\mu \rangle$) 
but also contain terms like 
$\mathcal{G}^{BC}_{\mu\nu} = \langle \mathcal A^B_\mu \mathcal A^C_\nu \rangle$. 
Starting with Eq. \eqref{sec3-130} one can generate an operator differential equation for the product 
$\widehat {\mathcal A}^B_\mu \widehat {\mathcal A}^C_\nu$ thus allowing the determination of the Green function $\mathcal{G}^{BC}_{\mu\nu}$
\begin{equation}
  \left\langle Q \left|
  \widehat {\mathcal A}^B(x) \partial_{y\nu} \widehat {\mathcal F}^{B\mu\nu}(x)
  \right| Q \right\rangle = 0.
\label{sec3-140}
\end{equation}
However this equation will in turn contain other, higher order Green functions. Repeating these steps leads to an infinite set of equations connecting Green functions of ever increasing order. This construction, leading to an infinite set of coupled, differential equations, does not have an exact, analytical solution and so must be handled using some approximation. \emph{The problem is the same as in the turbulent theory.} 

For obtaining a tube-like solution in SU(3) gauge theory we do similar assumptions as in Section \ref{turbulence} (here we follow to \cite{Dzhunushaliev:2006di}):
\begin{itemize}
	\item The gauge field components $A^a_\mu \in SU(2)$ belonging 
	to the small subgroup $SU(2)$ are in an ordered phase:
	\begin{equation}
	  \left\langle a^a_\mu (x) \right\rangle = \bar a^a _{\mu} (x).
	\label{sec3-150}
	\end{equation}
	$a^a_\mu$ is the analog of $\bar v_z$. 
	\item The gauge field components $A^m_\mu \in SU(3)/ SU(2)$) are in a disordered phase:
\begin{equation}
  \left\langle A^m_\mu (x) \right\rangle = 0.
\label{sec3-160}
\end{equation}
	\item The 2-point Green function $G_2(x, y)$ approximately are  
\begin{equation}
  G_2(x, y) = \left\langle A^m_\mu (x) A^n_\nu (y) \right\rangle = 
  - \eta_{\mu \nu} 
  f^{mpa} f^{npb} \phi^a(x) \phi^b(y), \quad \mu, \nu = 1,2,3 ,
\label{sec3-170}
\end{equation}
here $\phi^a$ is a real triplet scalar fields. They are the analog of fluctuating velocity $v_\rho$. 
	\item Schematicaly the 4-points Green 
	function $G_4(x, y, z, u)$ can be decomposed by following way
	\begin{eqnarray}
  	G_4(x, y, z, u) &=& \left\langle A^m_\mu(x) A^n_\nu(y) 
  	A^p_\alpha(z) A^q_\beta(u) \right\rangle = 
  	\left[G_2(x, y) G_2(z, u) + \text{ permutations by } (x,y,z,u) \right] + 
\nonumber \\
	&& 
	\left[ \delta^{mn} \eta_{\mu \nu} G_2(z, u) + 
	\text{ permutations by } (x,y,z,u) \right].
	\label{sec3-190}
	\end{eqnarray}
\end{itemize}
After some calculations one can reduce initial $SU(3)$ Lagrangian to 
\begin{equation}
\begin{split}
	\mathcal L_{eff} = 
  - \frac{1}{4} \left\langle \mathcal{F}^A_{\mu\nu} 
  \mathcal{F}^{A\mu\nu} \right\rangle = &
  - \frac{1}{4} F^a_{\mu\nu} F^{a\mu\nu} + 
  \frac{1}{2} \left( D_\mu \phi^a \right) 
  \left( D^\mu \phi^a \right) - 
  \frac{\alpha}{2} A^b_\mu A^{c \mu} \phi^c \phi^b - 
  \\
  &
  \frac{\lambda}{4} \left( \phi^a \phi^a - m^2 \right)^2 - 
   \frac{1}{2} \left( M^2 \right)^{\mu \nu} A^a_\mu A^a_\nu 
\end{split}
\label{sec3-200}
\end{equation}
where $\alpha, \lambda, m, \left( M^2 \right)^{\mu \nu}$ are some parameters. The details can be found in Ref. \cite{Dzhunushaliev:2006di}. Corresponding field equations are 
\begin{eqnarray}
  D_\nu F^{a\mu\nu} &=& g \epsilon^{abc} \phi^b
  D^\mu \phi^c - \left( M^2 \right)^{\mu \nu} A^a_\nu ,
\label{sec3-210}\\
  D_\mu D^\mu \phi^a &=& -\lambda \phi^a
  \left(
  \phi^b \phi^b - \mu^2
  \right) - \alpha A^b_\mu A^{a \mu} \phi^b
\label{sec3-220}
\end{eqnarray}
where $D_\mu = \partial_\mu + g \epsilon^{abc} A^b_\mu$ is the covariant derivative. Using following ans\"atz for Eq's \eqref{sec3-210} \eqref{sec3-220} 
\begin{equation}
\begin{split}
    &A^1_t(\rho) = \frac{f(\rho)}{g} ; \quad A^2_z(\rho) = \frac{v(\rho)}{g} ;
    \quad \phi^3(\rho) = \frac{\phi(\rho)}{g}; 
\\
		&
    \left( M^2 \right)^{\mu \nu} =\mathrm{diag} 
    \left\{ M_0, M_1, 0, 0 \right\}
\label{sec3-230}
\end{split}
\end{equation}
gives us
\begin{eqnarray}
    v'' + \frac{v'}{z} &=& v \left( \phi^2 - f^2 - M^2_1 \right),
\label{sec3-240}\\
    f'' + \frac{f'}{z} &=& f \left( \phi^2 + v^2 - M^2_0 \right),
\label{sec3-250}\\
    \phi'' + \frac{\phi'}{z} &=& \phi \left[ - f^2 + v^2
    	+ \lambda \left( \phi^2 - m^2 \right) 
    \right].
\label{sec3-260}
\end{eqnarray}
The numerical investigation of the field equations shows that there exists a regular cylindrically symmetric solution with nonzero color electric fields 
\begin{equation}
  E^3_z = \frac{1}{g} fv, \quad E^1_\rho = - \frac{1}{g} f'.
\label{sec3-270}
\end{equation}
The flux of the chromoelectric field $E^3_z$ is nonzero 
\begin{equation}
	\Phi_E = 2 \pi \int \limits_0^\infty E^3_z \rho d \rho  = 
	2 \pi \int \limits_0^\infty f v \rho d \rho  \neq 0
\label{sec3-280}
\end{equation}
which is similar to nonzero flux of a turbulent fluid across a core in Section \ref{turbulence}
\begin{equation}
	2 \pi \int \limits_0^\infty \rho_m \left(
		\bar v_z - v_\infty  \right)
	\rho d \rho \neq 0.
\label{sec3-290}
\end{equation}

\section{Conclusions and discussion}

From author point of view the problems of turbulence and quantum chromodynamics are connected with an infinite equations set connecting all correlation (for turbulence) or Green (for quantum chromodynamics) functions. This equations set can not be solved exactly. Heisenberg \cite{heisenberg} wrote that there is only one way to solve such equations system: to cut off the set using some assumption on a Green function and to decompose some Green function as the combination of Green functions of lower orders. This is exactly what we have presented in this paper. 

Now we would like to enumerate common features being used here for finding tube-like solutions presented above:
\begin{itemize}
 \item For cutting off an infinite equations set for correlation/Green functions the correlation/Green function of third/fourth order is decomposed via lower correlation/Green functions.
 \item All correlation functions are expressed via one function. The Green functions are expressed via two set of scalar fields. 
\end{itemize}
It is necessary to emphasize that between the turbulence and quantum chromodynamics there is a  difference. Any field theory has Lagrangian and field equations are corresponding Euler equations for this Lagrangian. The Navier-Stokes equtions are not Euler equations in the consequence of the presence of viscocity.

\section{Acknowledgments} 
I am grateful to the Research Group Linkage Program of the Alexander von Humboldt Foundation for financial support, to J. Kunz for invitation to Universit\"at Oldenburg for research.


\begin{thebibliography}{99}

\bibitem{Energy}
National Energy Research Scientific Computing Center 2004 Annual Report: Advances in Computational Science, 
http://www.nersc.gov/news/annual$\_$reports/annrep04/html/adv-comp-sci/03-whorled.html

\bibitem{wiki}
Wikipedia, 
http://en.wikipedia.org/wiki/Color$\_$confinement

\bibitem{hooft}
G.~'t~Hooft, in {\it High Energy Physics}, ed. A. Zichichi,
EPS International Conference, Palermo (1975); \\
S.~Mandelstam, {Phys.\ Rept.}  {\bf 23}, 245 (1976); \\
G.~'t Hooft,
Nucl.\ Phys.\ B {\bf 190}, 455 (1981).

\bibitem{kondo} 
Kei-Ichi Kondo, 
Phys. Rev. \textbf{D57}, 7467 (1998).

\bibitem{Chernodub:2005jh}
M.~N.~Chernodub,
Phys.\ Lett.\  B {\bf 637}, 128 (2006);
hep-th/0506107.

\bibitem{niemi}
A.~J.~Niemi,
JHEP {\bf 0408}, 035 (2004);\\
A.~J.~Niemi and N.~R.~Walet,
hep-ph/0504034.

\bibitem{faddeev}
L.~D.~Faddeev and A.~J.~Niemi,
Phys.\ Lett.\ B {\bf 525}, 195 (2002).

\bibitem{Balachandran:2005ev}
  A.~P.~Balachandran, S.~Digal and T.~Matsuura,
  Phys.\ Rev.\  D {\bf 73}, 074009 (2006)
  [arXiv:hep-ph/0509276].

\bibitem{Nakano:2007dr}
  E.~Nakano, M.~Nitta and T.~Matsuura,
  Phys.\ Rev.\  D {\bf 78}, 045002 (2008)
  [arXiv:0708.4096 [hep-ph]].

\bibitem{Eto:2009kg}
  M.~Eto and M.~Nitta,
  arXiv:0907.1278 [hep-ph].

\bibitem{Tong:2005un}
  D.~Tong,
  arXiv:hep-th/0509216.

\bibitem{Eto:2006pg}
  M.~Eto, Y.~Isozumi, M.~Nitta, K.~Ohashi and N.~Sakai,
  J.\ Phys.\ A  {\bf 39}, R315 (2006)
  [arXiv:hep-th/0602170].

\bibitem{Shifman:2007ce}
  M.~Shifman and A.~Yung,
  Rev.\ Mod.\ Phys.\  {\bf 79}, 1139 (2007)
  [arXiv:hep-th/0703267].

\bibitem{Tong:2008qd}
  D.~Tong,
  Annals Phys.\  {\bf 324}, 30 (2009)
  [arXiv:0809.5060 [hep-th]].

\bibitem{highTc}
G.~Baskaran and P.~W.~Anderson,
Phys.\ Rev.\ B {\bf 37}, 580 (1988);\\
P.~W.~Anderson, Science 235, 1196 (1987);\\
G.~Baskaran, Z.~Zou, and P.~W.~Anderson, Solid State Comm. 63, 973 (1987);\\
P.~A.~Lee, N.~Nagaosa, X.-G.~Wen,
e-print cond-mat/0410445, submitted to Rev. Mod. Phys (2004).

\bibitem{Dzhunushaliev:2002xr}
V.~Dzhunushaliev and D.~Singleton,
Mod.\ Phys.\ Lett.\ A {\bf 18}, 955 (2003); 
hep-th/0210287.

\bibitem{heisenberg}
W. Heisenberg, \textit{Introduction to the unified field theory of elementary particles.},
(Max - Planck - Institut f\"ur Physik und Astrophysik, Interscience
Publishers London, New York-Sydney, 1966).

\bibitem{Dzhunushaliev:2006di}
V.~Dzhunushaliev,
Science Echoes, Vol.4, No:3, 82-112 (2008);
hep-ph/0605070.


\end{thebibliography}
\end{document}